\documentclass[prb,preprint]{revtex4-1}
\usepackage{amsmath,amsfonts,graphicx,color}

\begin{document}

\title{Analysis of the circular track experiment measuring the one-way speed of light}
\author{Evan John Philip}
\affiliation{School of Physical Sciences, National Institute of Science Education and Research, Bhubaneswar 751005, India}
\email{evanphilip@gmail.com}
\date{\today} 
\begin{abstract}
All experiments attempting to verify the invariance of speed of light directly are based on two-way speed measurement. The challenge in one-way speed measurement, the requirement of spatially separated synchronised clocks, can be possibly circumvented by measuring the speed of light travelling in a closed path. An apparent violation of the invariance principle has been recently reported in the first experiment attempting to measure the one-way speed of light utilising this concept. This experiment is reanalysed here. It is found that the results of the experiment can be explained within the framework of relativity, without requiring any violation of the invariance principle.
\end{abstract}

\maketitle

\section{Introduction}
The invariance of the speed of light for all inertial observers is a basic postulate of the special theory of relativity.~\cite{b3} However, this postulate has not been experimentally established directly. It has only been experimentally established that the two-way speed of light is independent of the velocity of the inertial frame in which it is measured.~\cite{b4}

One of the major challenges in performing direct measurement of the speed of light during one-way transmission is the requirement of spatially separated, synchronised clocks. A possible way of circumventing this challenge is measuring the speed of light travelling in a closed path. This will enable the use of the same clock for the initial and final measurement, thereby eliminating the need for synchronisation.

The speed of light in an experiment based on this idea, conceived and performed by C. S. Unnikrishnan, was found to have an apparent dependence on the velocity of the observer.~\cite{b1} We present a reanalysis of this experiment and show that the experimental results can be explained within the framework of special relativity, without requiring any violation of the principle of invariance of speed of light.

\section{The experiment}
A brief description of Unnikrishnan's experiment is given below to facilitate the understanding of this paper.

The experiment is presented by the author in Ref.~\citenum{b2} as an extension of the simple idea of measuring the speed of light  by sending a light pulse between two synchronised clocks.
\begin{figure}[h]
\centering
\includegraphics[width=0.5\textwidth]{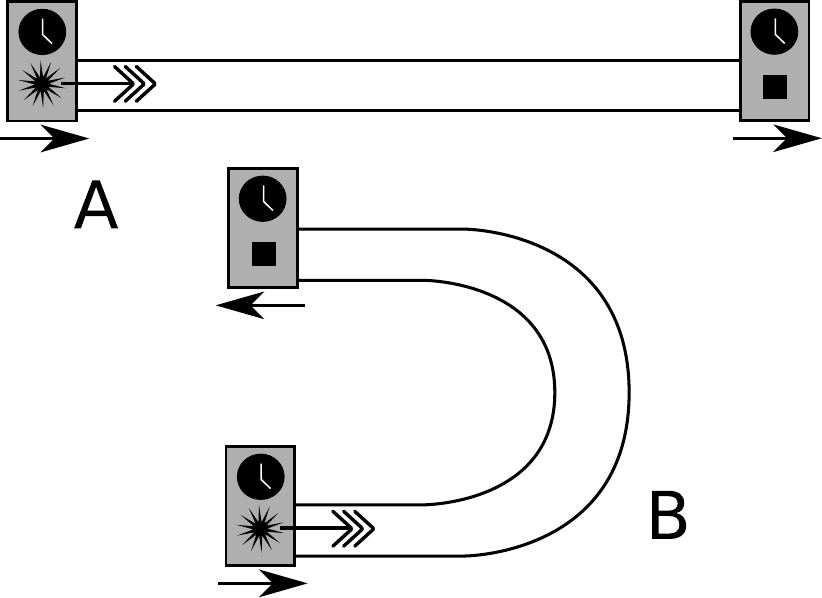}
\caption{(A) Simple setup for measuring of the speed of light in an inertial frame. (B) Modified setup.~\cite{b2}}
\label{fig:idea}
\end{figure}
According to the invariance principle, the measured speed of light must be independent of the constant velocity the two clocks may be moving with.
It was argued that, even if the path followed by the light pulse is not straight as in Figure~\ref{fig:idea} A, if the clocks are moving with constant velocities maintaing the length of the path, like in Figure~\ref{fig:idea} B, the speed of light must be invariant.~\cite{b2}

\begin{figure}[h]
\centering
\includegraphics[width=0.5\textwidth]{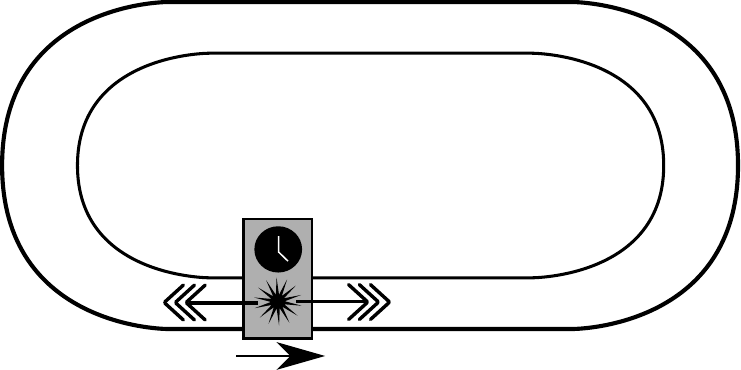}
\caption{Schematic diagram of the concept of the experiment.~\cite{b2}}
\label{fig:schematic}
\end{figure}

The path may be bent to make the initial and final clock the same, like in Figure~\ref{fig:schematic}. This maintains the total path length and eliminates the concern about synchronisation.

The invariance principle was to be validated by comparing the speeds of two light pulses, travelling in opposite directions, in a closed path, as schematically shown in Figure~\ref{fig:schematic}. The two light pulses were sent in opposite directions from an observer moving with constant velocity and the difference in their time of arrival was measured. Since the speed of the light pulses travelling in either direction must be the same with respect to the observer moving with constant velocity,  it was argued in Ref.~\citenum{b1} that the light pulses must arrive simultaneously. Note that the contribution of length contraction is the same in case of both pulses.

\begin{figure}[h]
\centering
\includegraphics[width=0.5\textwidth]{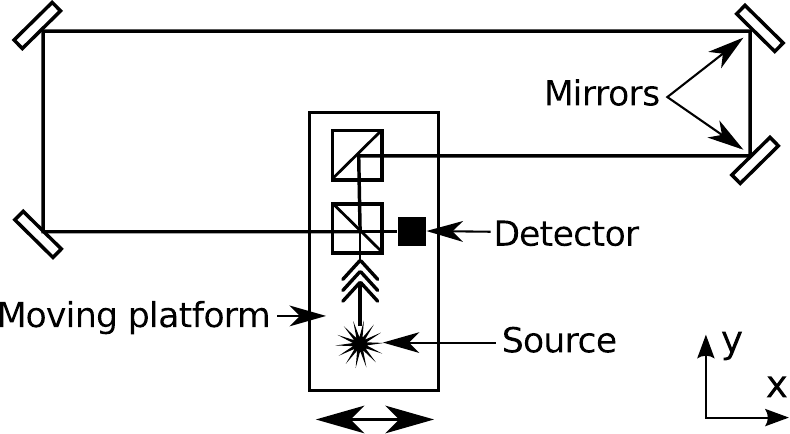}
\caption{Schematic diagram of the experimental setup used in Ref.~\citenum{b1}}
\label{fig:setup}
\end{figure}

A schematic diagram of the actual experimental setup used by Unnikrishnan is shown in Figure~\ref{fig:setup}.  
The detector interferometrically measures the difference in the time of arrival with a temporal resolution exceeding $10^{-18}$~s. The platform holding the detector is capable of moving with uniform velocity up to the order of $10^{-1}$~ms$^{-1}$. 

The result obtained by Unnikrishnan for a round trip length of the order of $2$~m can be found in Ref.~\citenum{b1}. It shows a clear linear dependence between the arrival time difference and the velocity of the observer. It was argued in Ref.~\citenum{b1} that this observation contradicts the invariance principle.

\section{Explanation of the Results}
\begin{figure}[h]
\centering
\includegraphics[height=0.35\paperheight]{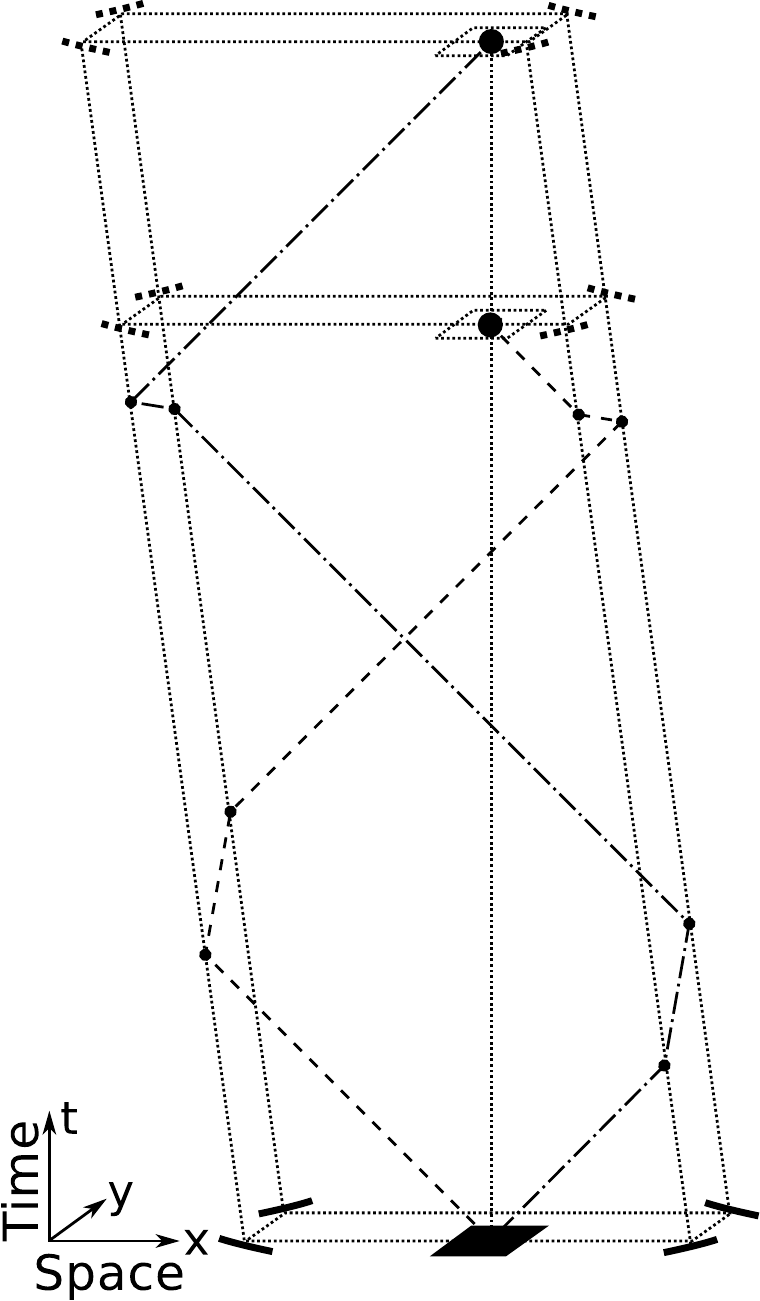}
\caption{Minkowski space-time diagram of the experiment from the frame of reference of the observer (detector).~\cite{b5} The mirrors appear to move in the opposite direction from this frame. The inclination of the world lines followed by light to the horizontal plane does not change due to the constant speed of light. The world lines followed by the mirrors are at a greater angle due to their lower speed (i.e. they fall within the light cone).}
\label{fig:Mkwsky}
\end{figure}

We will show below that this apparent violation of the postulate of invariance of speed of light arises because, according to the special theory of relativity, from the frame of reference of the inertial observer, the path travelled by the two light pulses are of unequal length. This can be seen easily by examining the Minkowski space-time diagram in Figure~\ref{fig:Mkwsky}.

If we consider the frame of reference of the observer, the mirrors keep moving while the light pulses travel from one mirror to another. From the start of the pulse to the detection of the pulse, this results in a change in the effective distance between the mirrors. Contrary to naive expectation, according to the special theory of relativity, the total distance travelled by the two light pulses is seen to be unequal (see Figure~\ref{fig:Mkwsky}). It turns out that it was erroneous to expect A and B of Figure~\ref{fig:idea} to yield the same result.

For the purpose of comparison with experimental results, the difference in the time of arrival can be calculated. This can be accomplished by writing the equations for the instantaneous positions of the mirrors from the frame of reference of the observer, calculating the time for the light pulse to travel each segment separately and summing them, to obtain the total time of travel for each pulse. 

We are expecting an effect which is first order in velocity. Since the velocity of the inertial observer is much less than the speed of light, all contributions arising from higher orders of velocity can be ignored.

The difference between the time of arrival of the two light pulses at the detector is found to be
$$\Delta t \approx \frac{2l}{c^2} v$$ 
where $l$ is the length of the path, $v$ is the speed of the detector with respect to the rest frame of the mirrors and $c$ is the speed of light.

Substituting $l = 2$~m  in the above expression, we obtain
$$\Delta t = (4.44\times 10^{-17} s^2 m^{-1}) v$$
The slope calculated above is comparable to the slope of the experimental plot in Ref.~\citenum{b1}.

Figure~\ref{fig:Mkwsky} shows the basic difference between our analysis and the analysis in Ref.~\citenum{b1}. In our analysis, with respect to the observer (detector), the world lines of the mirrors are inclined in the same direction, since all the mirrors have the same relative velocity with respect to the observer. In Ref.~\citenum{b1}, the world lines of the mirrors are shown to be inclined in different directions. 

\section{Concluding Remarks}
We have reanalysed the experiment in Ref.~\citenum{b1} and have shown that the difference between the time of arrival of the two light pulses and its dependence on the velocity of the observer does not violate the principle of invariance of speed of light. As explained above, the time of traversals correspond to that of two different distances, so they cannot be expected to be the same. The velocity dependence of the difference in the time of arrival naturally arises as the difference in the distances of traversal depend on the velocity of the observer. The experimental results of Ref.~\citenum{b1} appear to be numerically in reasonable agreement with the calculations performed using the special theory of relativity, without requiring any violation of the principle of invariance of speed of light.

\begin{acknowledgments}
I would like to thank Prof. S. C. Phatak and Prof. A. M. Srivastava for encouraging me to prepare the manuscript and for their critical comments on it. I would also like to thank my friend Ankit Pandey for inspiring me to tackle the problem.
\end{acknowledgments}

\bibliographystyle{unsrtnat}
\bibliography{EvanBibliography}

\end{document}